\documentclass[aps,prl,amsmath,amssymb,10pt,twocolumn,superscriptaddress,notitlepage,doublespace]{revtex4-1}
\usepackage{graphicx}
\usepackage{dcolumn}% Align table columns on decimal point
\usepackage{bm}% bold math
\usepackage{amssymb}
\usepackage{braket}

\begin{document}

\title{Precision Measurement of Time-Reversal Symmetry Violation with Laser-Cooled Polyatomic Molecules}

\author{Ivan Kozyryev}
\email{ivan@cua.harvard.edu}
\affiliation{Department of Physics, Harvard University, Cambridge, MA 02138, USA}
\author{Nicholas R. Hutzler}
\email{hutzler@caltech.edu}
\affiliation{Division of Physics, Mathematics, and Astronomy, California Institute of Technology, Pasadena, CA 91125, USA}
\affiliation{Department of Physics, Harvard University, Cambridge, MA 02138, USA}
\date{\today}

\begin{abstract}
Precision searches for time-reversal symmetry violating interactions in polar molecules are extremely sensitive probes of high energy physics beyond the Standard Model.  To extend the reach of these probes into the PeV regime, long coherence times and large count rates are necessary.  Recent advances in laser cooling of polar molecules offer one important tool -- optical trapping.  However, the types of molecules that have been laser-cooled so far do not have the highly desirable combination of features for new physics searches, such as the ability to fully polarize and the existence of internal co-magnetometer states.  We show that by utilizing the internal degrees of freedom present only in molecules with at least three atoms, these features can be attained simultaneously with molecules that have simple structure and are amenable to laser cooling and trapping.
\end{abstract}
\maketitle

\newcommand{\E}{\mathcal{E}}
\newcommand{\B}{\mathcal{B}}

Precision measurements of heavy atomic and molecular systems have proven to be a powerful probe of high energy scales in the search for New Physics Beyond the Standard Model (BSM) \cite{DeMille2015}.  For example, the limit on the electron's electric dipole moment (EDM), set by the ACME collaboration using ThO, is sensitive to T-violating BSM physics at the $\gtrsim$ TeV scale \cite{Baron2014}.  This sensitivity relies on the ability to experimentally access the large effective electromagnetic fields ($>10$ GV/cm) present in heavy polar molecules by fully polarizing them in the laboratory frame.  This makes the experimental challenges of working with such a complex species worth the effort.

Despite the success of ACME, a current limitation of that experiment and all present molecular beam experiments is that their coherence time is limited to a few milliseconds by the beam transit time through an apparatus of reasonable size.  Since EDM sensitivity scales linearly with coherence time, trapping neutral molecules has the potential to increase sensitivity by many orders of magnitude.  Trapped molecular ions have shown great power in EDM searches \cite{Cairncross2017}, primarily due to their long coherence time of $\sim$1 s.  Neutral species offer the ability to increase the number of trapped molecules much more easily and essentially without limit compared to ions, while retaining strong robustness against systematic errors.  Here we show that laser-cooled and trapped polyatomic molecules offer a combination of features not available in other systems, including long lifetimes, robustness against systematic errors, and scalability, and present a feasible approach to access PeV-scale BSM physics.

A very promising route to trapping EDM-sensitive molecules is direct laser cooling and trapping from cryogenic buffer gas beams (CBGBs), which has advanced tremendously in the last few years \cite{Stuhl2008,Shuman2010,Hummon2012,Barry2014,Chae2017,Truppe2017Slow,Truppe2017SubDoppler,Anderegg2017}.  The molecules that have been cooled so far posses an electronic structure that makes them amenable to laser cooling, but also precludes the existence of $\Omega-$doublets, such as the $^3\Delta_1$ molecular state used in the two most sensitive electron EDM measurements \cite{Baron2014,Cairncross2017}.  These doublets enable full polarization and ``internal co-magnetometry,'' which allows for the reversal of the EDM interaction without changing any lab fields.  These features afford crucial robustness to systematic effects, especially as sensitivity continues to improve.  There are a number of diatomic molecules with good sensitivity to BSM physics that are laser-coolable, such as BaF \cite{Chen2016}, RaF \cite{Isaev2010},  and YbF \cite{Smallman2014}, though these molecules do not have closely spaced levels of opposite parity.  They therefore require large and technically challenging lab electric fields $\gtrsim 10$ kV/cm  in order to be sensitive to the EDM, cannot be fully polarized, and do not admit internal co-magnetometers -- all of which leave them vulnerable to challenging systematic effects.  Combining the requirement of laser cooling with the requirement of full polarization and internal co-magnetometers eliminates all known choices of diatomic molecules.   RaOH, a laser-coolable polyatomic molecule with BSM physics sensitivity, was previously considered for a precision measurement in the ground vibrational state \cite{Isaev2016RaOH}, meaning that it would still suffer from the same drawbacks as diatomics.

 We show here that low-lying excited vibrational modes in polyatomic molecules, which have not been previously considered for precision measurements, allow full polarization and internal co-magnetometry via generic degrees of freedom, and are excellent candidates for a new class of precision measurements.   Degenerate bending modes in these states give rise to lab-accessible angular momentum with a projection along the molecular dipole, enabling full polarization in small fields analogous to $\Omega-$doublets.  However, unlike in $\Omega-$doublets these degrees of freedom are not coupled to the electronic spin and therefore do not interfere with either laser cooling properties or sensitivity to BSM physics.  These structures are generic, and can be used to access these advantages with any atom that is sensitive to BSM physics.
  
The molecules we will consider consist of an alkaline earth (or alkaline earth-like) atom monovalently and ionically bonded to some functional group.  However, the ideas discussed are generally applicable to other polyatomic species.  We show that these molecules have the significant additional advantage of being laser-coolable, as was recently demonstrated with the polyatomic molecule SrOH \cite{Kozyryev2017SrOH} and proposed for a number of other species \cite{Isaev2016Poly,Kozyryev2016Poly,Isaev2016RaOH}.  The essential property is the non-bonding $s$ electrons being removed from the bonding region by orbital hybridization \cite{Ellis2001}, resulting in highly diagonal Franck-Condon factors (FCFs).  This property is not strongly dependent on the type of functional group bound to the metal atom \cite{Isaev2016Poly,Kozyryev2016Poly}.  Thus, polyatomic molecules isoelectronic to suitable diatomic candidates for fundamental physics searches such as BaF, YbF, HgF, and RaF, have promise for laser cooling. Since the BSM physics sensitivity also comes from the non-bonding electron, it is largely independent of the bonding partners \cite{Isaev2016RaOH}.  Furthermore, these polyatomic molecules are readily created in molecular beams and have well-studied and understood spectra \cite{Ellis2001}.

%\subsection{Linear molecules}

\begin{figure}[htbp]
	\centering
		\includegraphics[width=0.40\textwidth]{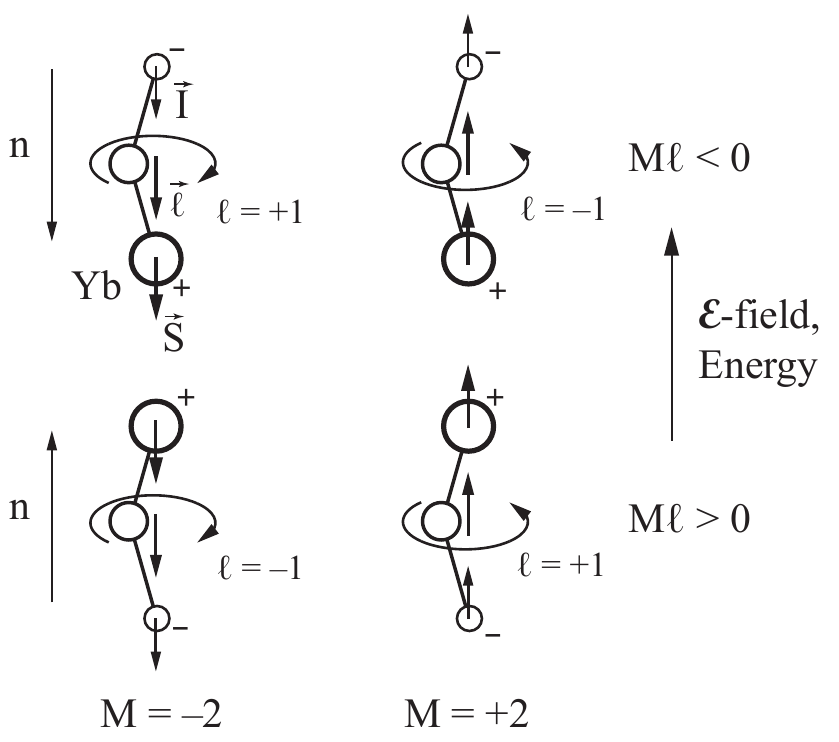}
	\caption{Angular momentum in the fully-polarized stretched states $F=|M|=2$, in which each of the component angular momenta are stretched as well.  The quantum numbers $S, \ell$, and $I$ are indicated at the top-left.  The internuclear axis points from the negative to the positive ion, meaning that the dipole moment lies along the internuclear axis.  Compare the very similar figure for a $^3\Delta_1$ state like WC $X$ \cite{Lee2009} or ThO $H$ \cite{Baron2016}.  Like $\Omega$, $\ell$ is quantized in the molecule frame, which is why the direction of the vector $\vec{\ell}$ on the figure and the value of $\ell$ may differ.  Since the EDM shift is $\propto\vec{S}\cdot n\propto\ell$, we can see that this interaction is reversed in the upper/lower Stark shifted states.}
	\label{fig:Dumbbell}
\end{figure}

We will consider linear and symmetric top molecules, starting with the simplest type of molecule with the required characteristics -- a linear non-symmetric triatomic, $XYZ$.  There are three distinct vibrational modes in this molecule \footnote{The numbers $\nu_1,\nu_2,\nu_3$ also denote symmetric stretch, bend, and asymmetric stretch respectively \cite{Townes1955}.  The notation used in the text is valid for molecules with atomic masses $m_X\gg m_Y\gg m_Z$, like those considered here \cite{Oberlander1995}}: $X-Y$ stretch, bend, and $Y-Z$ stretch, denoted by vibrational quantum numbers $(\nu_1,\nu_2,\nu_3)$ respectively.  The $\nu_2$ mode is doubly-degenerate, as the bending can occur in two perpendicular directions.  Since the molecule is symmetric about its axis, the eigenstates are sums of these two motions and the molecule has angular momentum $\ell$ along its symmetry axis, as shown in figure \ref{fig:Dumbbell}.  In the excited $\nu_2=1$ mode, there are two such states with $\ell=\pm 1$, denoted $\nu_2^{\pm\ell}$.  Analogous to $\Omega-$doubling, Coriolis interactions lift the degeneracy between the even and odd parity states $\ket{1^{+1}}\pm\ket{1^{-1}}$, resulting in a parity doublet of size $q\sim\mathcal{O}(B_e^2/\omega_2)$, where $B_e$ is the rotational constant and $\hbar\omega_2$ is the vibrational energy for this mode \cite{Herzberg1942}.  For the types of species we will consider this splitting is typically $\sim 10$ MHz, and can therefore be mixed in moderate lab fields of $\sim 100$ V/cm.  The resulting polarized states are suitable to search for T-violating physics, and are such a generic feature that we can find them for polyatomics with any desired heavy atom.

As a specific example, we consider an electron EDM search in YbOH.  We choose this molecule as our example case because it is readily created in a molecular beam, has been studied spectroscopically \cite{Melville2001, Melville2001Thesis}, is sensitive to many T-violating effects such as the electron EDM \cite{Hudson2011} and nuclear magnetic quadrupole moment \cite{Flambaum2014} via the heavy Yb atom, and is a suitable candidate for direct laser cooling as we shall describe later.  We stress again that the presented results do not depend on the specific properties of YbOH, and are quite generic.  This molecule has a $^2\Sigma$ electronic ground state arising from a Yb-centered electron spin $S=1/2$.  $S$ couples to the combined, total rotational and vibrational angular momentum $N$ via spin-rotation $\gamma N\cdot S$ to form $J=N+S$.  The H nucleus has spin $I=1/2$, which couples to $J$ via Fermi contact $bS\cdot I$ to form the total angular momentum $F=J+I$, with projection $M$ on the lab $z-$axis.   A schematic of these angular momenta is shown in figure \ref{fig:Dumbbell}, and the structure is discussed further in the appendix.  This is highly analogous to similar $^2\Sigma$ electronic states in diatomic molecules, with the important difference that $N$ includes $\ell$, a quantum number absent in diatomics.  

Consider the $\nu_2=1$ state, which lies above the absolute ground state by about 300 cm$^{-1}\approx $10 THz, and has an $\ell-$doubling constant of $q\approx -10$ MHz, a spin-rotation constant $\gamma\approx 30$ MHz, and a hyperfine constant $b\approx 2$ MHz.   The lifetime of this low-lying state is estimated to be $\gtrsim 10$ s in the appendix.  To prove that this state is a good candidate for an EDM search, we will examine its Stark, Zeeman, and EDM shifts.

\begin{figure}[htbp]
	\centering
		\includegraphics[width=0.45\textwidth]{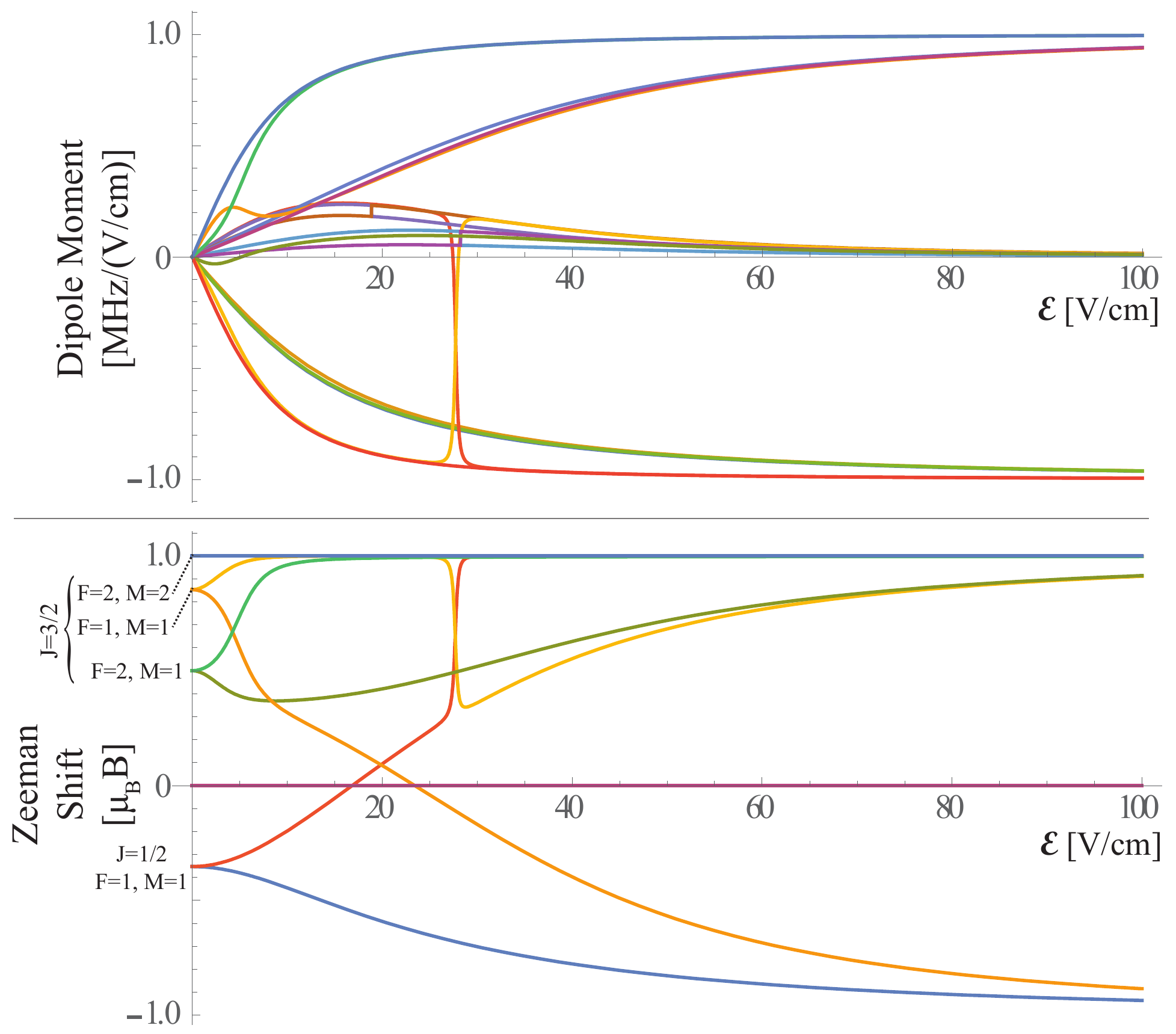}
	\caption{Electric dipole moments (top) and Zeeman shifts (bottom) with $\gamma=$30 MHz, $q=-10$ MHz, $b=2$ MHz, and $d=4$ D, representative of YbOH.  The Zeeman shift is in units of $\mu_B\B$ in a small magnetic field.  The dipole moment is also signed polarization, as described in the text.  The jumps indicate avoided crossings.  The labels on the left side correspond to the zero-field eigenstates.  The colors in both plots correspond to the same states.}
	\label{fig:DipoleMoments}
\end{figure}

Consider an electric field $\E$ applied along the lab $z$ axis, and assume a (typical) dipole moment of $d=4$ D, which saturates to a Stark shift of 1 MHz/(V/cm).  This means that the dipole moment in these units is also the signed polarization, both of which are shown in figure \ref{fig:DipoleMoments}.  These levels were calculated by diagonalizing the $N=1$ states including the Stark, spin-rotation, Fermi contact hyperfine, and $\ell-$doubling interactions as described in the appendix.  We consider $\E$ small enough to neglect contributions from $N=2$.

YbOH has states with $>90\%$ polarization at fields of $\sim 40$ V/cm, and $>99.9\%$ at 250 V/cm.  Since the EDM shift is proportional to the polarization, this means that we can easily saturate the EDM sensitivity in the lab frame.  The states with the largest polarizability are the stretched $F=|M|=2$ states, which admit a simple intuitive diagram of angular momentum orientation, shown in figure \ref{fig:Dumbbell}.

We now consider a small magnetic field $\B$ parallel to $\E$, and calculate the combined Stark and Zeeman shifts. Figure \ref{fig:Dumbbell} suggests that these polarized states have a linear Zeeman shift (electron spin either aligned or anti-aligned with $\B$ depending on the sign of $M$), which is confirmed by diagonalizing the full Hamiltonian (see appendix.)  The Zeeman shifts in a small magnetic field as a function of applied electric field are shown in figure \ref{fig:DipoleMoments}.

There are electric fields where the effective $g-$factors cross zero.  Unlike cases where this has been considered previously \cite{Shafer-Ray2006}, these fields are quite small.  Unfortunately these states should have little EDM-sensitivity; zero $g-$factor means that the electron is not oriented in the lab, and since there is no strong coupling of the electron spin to the molecular internal frame,  the electron cannot be aligned in the molecule frame either.  However, these states could be very useful for systematic checks of $\E-$field dependence of spin precession without a background signal due to the much larger Zeeman effect.

Now consider the sensitivity to the electron EDM.  Both $S$ and $\hat{n} \equiv sign(M\ell)$ (the molecule dipole moment orientation) are stretched and aligned along the lab $z-$axis, so the EDM shift in the polarized limit is simply given by 
$\Delta_{EDM} \propto S\cdot n \propto  sign(S\cdot z)sign(n\cdot z) = \ell$, perfectly analogous to the shift $\Delta_{EDM}\propto\Omega$ for a fully-polarized diatomic molecule in a state with $\Omega-$doublets.  The EDM shift reverses sign upon changing the molecule orientation, which provides the desired internal co-magnetometer via spectroscopic reversal.

The stretched states have the simplest interpretation, but other states are equally useful. In particular, for both the Stark and Zeeman effects all of the states saturate to either the same absolute value, or zero.  For the Stark effect, this is simple to understand; only $N$ has any interaction with the applied field to first order, so $N=1$ should have at most three values of dipole moment in the fully-polarized limit.  The Zeeman shift saturates as a result of the applied electric field decoupling the molecular dipole moment and symmetry axis from the electron spin and occurs when $|d\E|\gtrsim|\gamma|$, analogous to the decoupling of atomic electron and nuclear spins in a high magnetic field.  The symmetry axis and electron spin are aligned in the lab for any Stark-shifted state with $M\neq 0$, meaning that the EDM sensitivity saturates to the same value for any pair of $\pm M$ states in the Stark-shifted manifolds.  This means that we can use any pair of $\pm M\neq 0$ states (with the same Stark shift) to perform the measurement, eliminating the need for potentially difficult coherent preparation of states with large angular momentum projection difference.  Note that all such states have $>99\%$ polarization in a 300 V/cm field.

%\subsection{Laser Cooling}

\begin{figure}
\begin{centering}
\includegraphics[width=0.4\textwidth]{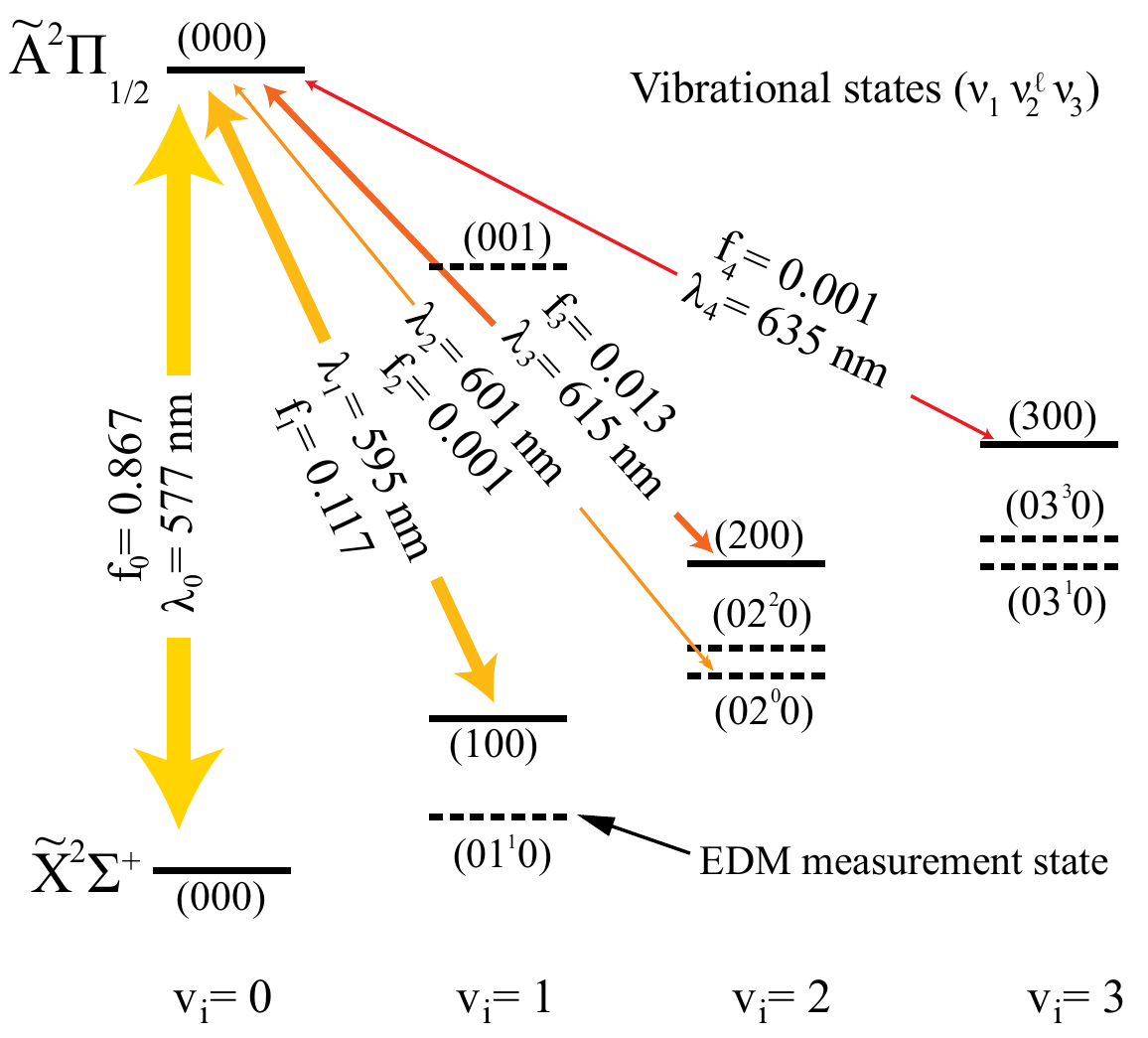}\protect\caption{\label{fig:Proposed-photon-cycling-scheme}Proposed photon cycling
and laser cooling scheme for YbOH. Thicker lines indicate stronger
transitions with appropriate Franck-Condon factors ($f$) and wavelengths
($\lambda$) indicated. The energy splittings are not to scale. }
\par\end{centering}
\end{figure}

Now we shall discuss how these molecules can be laser cooled, and show that it can be performed efficiently.  This is a necessary step for loading a magneto-optical trap (MOT), which is a very promising step in the path to trapping with long coherence times.
Laser cooling and trapping of YbOH is feasible using the scheme originally proposed for CaOH \cite{Isaev2016Poly} and experimentally
demonstrated with SrOH \cite{Kozyryev2017SrOH}. Like SrOH, YbOH is an ionic molecule
with the two lowest electronic states $\tilde{X}^{2}\Sigma^{+}$ and
$\tilde{A}^{2}\Pi$ originating primarily from $4f^{14}6s\sigma$
and $4f^{14}6p\pi$ Yb\textsuperscript{+} atomic orbitals, respectively. Figure \ref{fig:Proposed-photon-cycling-scheme} shows the main $\tilde{X}^{2}\Sigma^{+}\left(000\right)\leftrightarrow\tilde{A}^{2}\Pi_{1/2}\left(000\right)$
laser cooling transition $\lambda_{0}$ as well as the dominant off-diagonal
vibrational decay channels in the Born-Oppenheimer (BO) approximation with FCFs $f\gtrsim0.001$. With four repumping lasers $\lambda_{1-4}$, shown in figure \ref{fig:Proposed-photon-cycling-scheme}, we can scatter thousands of photons. This allows for transverse beam compression
via the Doppler force leading to at least an order of magnitude enhancement
in on-axis peak beam density \cite{DeMille2013}, directly resulting in enhanced MOT loading \footnote{Here we assume experimental geometry similar to precision measurement
EDM experiments like ACME \cite{Baron2014} or magneto-optical trapping experiments
for diatomic molecules \cite{Barry2014} and consider cooling in 2D.}. Efficient 1D Sisyphus laser cooling of triatomic molecules has been
demonstrated with only a few hundred photons \cite{Kozyryev2017SrOH} and upon 2D implementation
in YbOH will lead to $\times6$ increased flux for MOT loading. Scattering of $\gtrsim10^4$ photons per molecule should be possible with five vibrational repumpers, enabling longitudinal slowing \cite{Barry2012} and direct magneto-optical trapping \cite{Barry2014}.

CBGBs can be used to produce high brightness and low velocity beams of nearly any small molecule \cite{Hutzler2012}.  Many molecules of the type under consideration, for example YbOH \cite{Melville2001} and YbCCH \cite{Loock1997}, have been created in beams by ablating metal into an inert carrier gas mixed with a reactive gas like H$_2$O$_2$ and HCCH respectively, a technique commonly implemented in CBGBs as well.

While excited bending vibrations are populated during the laser ablation
process \cite{beardah1999}, they quench relatively quickly through inelastic
collisions with helium buffer gas \cite{kozyryev2015}. A CBGB of YbOH will
mostly include molecules in the lowest vibrational state $\left(000\right)$, and will require pumping into the excited bending mode.  This pumping can be achieved via the excited $\mu^2\Sigma^{(+)}$ state, as discussed in the appendix.

Linear triatomic molecules are the easiest to understand, but slightly more complex molecules offer a possible advantage.  In particular, for symmetric top molecules the $K-$doublet is analogous to the $\ell-$doublet, where $K$ is the projection of the total rotational, orbital, and vibrational angular momentum on the symmetry axis \cite{Townes1955}.  The advantageous features of $\ell-$doublets are preserved, as is the ability to laser-cool species such as YbCH\textsubscript{3} and YbOCH\textsubscript{3} \cite{Isaev2016Poly,Kozyryev2016Poly}. However, $K-$doublet splittings are even smaller, typically $\lesssim$ kHz, meaning that complete polarization requires only $\lesssim$ 1 V/cm electric fields, and the excited $K$ levels are even closer to the ground state (typically $\sim$ 100 GHz).  Other molecular structures may reveal additional advantages.

%\subsection{Summary and further directions}

$\ell-$ and $K-$ doublets are quite generic and not limited to monovalent alkaline earths. Species such as RaOH \cite{Isaev2016RaOH}, RaCO, TlOH, ThCH, LuCH, PbOH, HfCH, LuCO, and many more (both diamagnetic and paramagnetic) can be used to search for a wide array of BSM physics beyond the electron EDM, including nuclear magnetic quadrupole moments, nuclear EDMs, nuclear Schiff moments, parity violation, and so on.  Some of these molecules may not be as readily laser-cooled, though we could potentially create ``custom'' species with a laser-coolable atom, for example TaCOCa.  Such species also have the potential for optical-cycling readout on the ``BSM physics atom'' via coupling of different spin polarizations to various internal states involving the ``laser-cooling atom.''  Combining such laser-coolable centers would be advantageous even for species that can be laser-cooled directly; a molecule such as YbCCCa would offer increased scattering rates and optical forces, and even more internal co-magnetometry.  Since both YbCCH and CaCCH can be created in a beam by reactions of the metals with HCCH \cite{Loock1997,Li1996CaCCH}, there is a promising path to creating such molecules.  We can also consider molecules for ion trap experiments, where the internal co-magnetometers are necessary \cite{Cairncross2017} since there is no ability to reverse the applied electric field, such as LuOH$^+$ or RaOH$^+$.  Additionally, the combination of laser cooling, optical readout, and linear Stark shifts in small fields could be useful for quantum information processing and quantum simulation \cite{Micheli2006,Wei2011}.

%Additionally, our analysis of the combined Stark and Zeeman shifts for YbOH indicates the advantages of using laser cooled and trapped linear triatomic molecules in excited bending states for quantum computation processing and quantum simulation applications \cite{Micheli2006,Wei2011} with the ability for non-destructive optical readout.

As an example of what sort of gains are to be had with this approach, consider $10^6$ trapped molecules with 10 second coherence time, 50$\%$ preparation/detection efficiency, and one week of operation.  Such an experiment would increase sensitivity to the electron EDM by four orders of magnitude above the current limit, reaching into the PeV regime \cite{Baron2014,Engel2013}. 

In conclusion, we have analyzed an experimentally viable approach for measuring T-violating interactions with simple polyatomic molecules in order to search for BSM physics at the PeV scale. Linear and symmetric top molecules containing a heavy metal atom like Yb provide a robust platform for an EDM search via laser cooling and trapping, and are the first system to combine the primary advantages of the competing approaches.

\begin{acknowledgments}

We thank John M. Doyle for his enthusiasm about polyatomic molecules, which planted the ideas that led to this scheme, and for feedback on the manuscript.  We also thank Timur Isaev, Tim Steimle, David DeMille, and John M. Doyle for insightful discussions.

\end{acknowledgments}

\bibliography{library}
\newpage
\section{Appendix}

\subsection{Molecular structure and Stark/Zeeman shifts}

We shall examine the structure and Stark/Zeeman shifts of a triatomic molecular degenerate bending mode $\nu$ with vibrational angular momentum $\ell$ and energy $\omega$. Consider the case of a heavy atom at the end of the molecule with a light functional group, $\ell=1$, $S=I=1/2$, with the nuclear spin not on the heavy atom, such as YbOH.  The quantum numbers and couplings are $F=I+J$ and $J=N+S$.  $J$ includes the degenerate bending mode angular momentum $\ell$, therefore $N$ is a coupling of molecule rotation and $\ell$ and can take the values $N=|\ell|, |\ell|+1, |\ell|+2,\ldots$.  We shall consider the case $N=|\ell|=1$, and therefore often leave $S,I,$ and $N$ out of kets.  We shall only consider externals fields small enough that we can neglect mixing with the $N=2$ state.

\subsubsection{Zero-field}

\begin{figure}[htbp]
	\centering
		\includegraphics[width=0.45\textwidth]{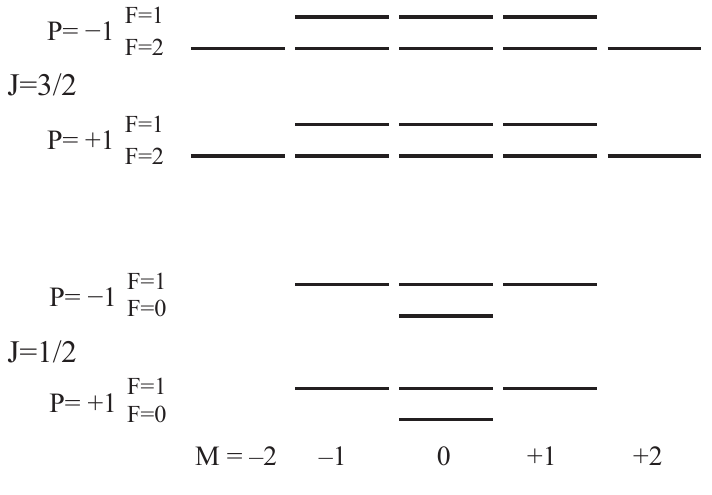}
	\caption{Structure of a $^2\Sigma$ electronic, $N=|\ell|=1,$ $S=I=1/2$ state in zero field. The scale (both relative and overall) and ordering of levels may not be accurate for every molecule, but are representative of YbOH.}
	\label{fig:ZeroFieldStructure}
\end{figure}

The most important interactions for us are spin-rotation, Fermi contact hyperfine, and $\ell-$doubling.  The energy level diagram in zero external fields is shown schematically in figure \ref{fig:ZeroFieldStructure}.

\emph{Spin-rotation:} In a diatomic $^2\Sigma$ molecule, the spin-rotation interaction $\gamma S\cdot N$ splits each $N$ level into 2 $J-$levels with $J=N\pm 1/2$.  The physical origin is the electron spin interacting with the magnetic field of the orbiting nucleus, so it might seem odd that in the ``non-rotating" $N=|\ell|=1$ state we get a similar interaction.  Since $\ell$ involves the rotation of a nucleus about the symmetry axis, this creates a magnetic field that interacts with the electron spin.  The magnitude of this effect is emperically similar to the usual spin-rotation effect, since the $\gamma$ constant in the $\nu=0$ and $\nu=1$ states are similar \cite{Fletcher1995}.  For YbOH, $\gamma\approx 30$ MHz \cite{Melville2001}.  One can also make a simple, semi-classical argument that the magnetic field created by the rotating nucleus in both cases is similar.

Brown and Carrington \cite{Brown2003} equation (9.89) gives the formula for the corresponding diatomic case, which therefore excludes $\ell$.  In \cite{Fletcher1995}, they find that the alkaline earth hydroxide spin-rotational structure is well-described by this Hund's case (b) Hamiltonian, so we use the diatomic formula with the addition that $\ell$ is not changed:
\begin{multline}
\braket{N\ell SJIFM | H_{SR} | N\ell SJIFM} = \\
\gamma \delta_{\ell\ell'}(-1)^{N+J+S} \begin{Bmatrix}S & N & J \\ N & S & 1\end{Bmatrix} \\
\times \left[S(S+1)(2S+1)N(N+1)(2N+1)\right]^{1/2}.
\end{multline}
This interaction is diagonal in our basis, as expected.

\emph{Hyperfine:}  The hyperfine interaction consists of a Fermi contact interaction $\beta_\eta S\cdot I$.  There is also a spin-spin interaction between $S$ and $I$ characterized by the constant $c$, but the matrix elements in this state suppress it to be much smaller than the Fermi contact, and we shall ignore it.  The matrix elements for Fermi contact can be found in Hirota \cite{Hirota1985} equation (2.3.80):
\begin{multline}
\braket{N'\ell'SJ'IF'M' | H_{HF} | N\ell SJIFM} = \\
b_\eta \delta_{MM'}\delta_{FF'}\delta_{NN'}\delta_{\ell\ell'} (-1)^{N+S+J'}(-1)^{J+I+F+1} \\
\times \left[S(S+1)(2S+1)I(I+1)(2I+1)\right]^{1/2} \\
\times\left[(2J'+1)(2J+1)\right]^{1/2} \begin{Bmatrix} I &J' & F \\ J & I & 1 \end{Bmatrix}\begin{Bmatrix} S & J' & N \\ J & S & 1 \end{Bmatrix}
\end{multline}
The constant $b_\eta$ is also written as $b_\eta = b-c/3$.  For proton hyperfine from an OH group, such as in YbOH, $b\approx c \approx b_\eta \approx$ 2 MHz \cite{Fletcher1993}.

\emph{$\ell-$doubling:}   Since $\ell$ breaks parity symmetry, the zero-field (parity, $P$) eigenstates are
\begin{equation}
\ket{\pm,J,F,M} = \frac{\ket{+\ell,J,F,M} \pm \ket{-\ell,J,F,M}}{\sqrt{2}}.
\end{equation} 
Because $\ell$ is the projection of $J$ onto the symmetry axis, it is very analogous to $\Omega$ for the case of a diatomic, and $K$ for the case of a symmetric top.   In fact, for a degenerate bending mode $\nu$, the state $\nu=\ell=1$ correlates to $K=1, \nu=0$ \cite{Herzberg1967}.  Coriolis interactions lift the degeneracy between the opposite-parity states $\pm$ and result in $\ell-$type doubling.  Each state $\ket{\pm,J,F,M}$ acquires an energy shift of $\pm \frac{1}{2}qN(N+1) = \pm q$.  The magnitude of this effect is generally the same order as the vibration-rotation interaction constant $\alpha$, defined as $B_\nu = B_e-\alpha(\nu+1)$, and is generally of order $\sim B_e^2/\omega$ \cite{Herzberg1942}.  For YbOH, we can use the value of $q\approx-10$ MHz from another heavy hydroxide BaOH, since the rotational constant and vibrational energy are similar \cite{Fletcher1995}.

\subsubsection{Stark effect}

We shall only consider $M\geq 0$, keeping in mind that the Stark shifts are even in $M$ by parity symmetry.  The matrix elements of the Stark Hamiltonian $H_S$ of an electric field $\E$ in the lab $z-$direction are given by Hirota \cite{Hirota1985} equation (2.5.35):
\begin{multline}
\braket{N'\ell'SJ'IF'M'|H_S|N\ell SJIFM} = \\
\E d(-1)^{F'-M'}\begin{pmatrix}F' & 1 & F \\ -M' & 0 & M\end{pmatrix} \\
\times (-1)^{J'+I+F+1}[(2F'+1)(2F+1)]^{1/2}\begin{Bmatrix} J' & F' & I \\ F & J & 1\end{Bmatrix} \\
\times (-1)^{N'+S+J+1}[(2J'+1)(2J+1)]^{1/2}\begin{Bmatrix} N' & J' & S \\ J & N & 1\end{Bmatrix} \\
\times (-1)^{N'-\ell'}\begin{pmatrix}N' & 1 & N \\ -\ell' & 0 & \ell\end{pmatrix},
\end{multline}
where we have taken $p=q=0$, written $T^1_{p=0}(E)=\E$ and $T^1_{q=0}(d)=d$, and set $K\rightarrow\ell$.  Note that the first and last lines of this formula look just like a Hund's case (a) or (c), with the middle two lines coming from the couplings $F=I+J$ and $J=N+S$, respectively.

\begin{figure}[htbp]
	\centering
		\includegraphics[width=0.45\textwidth]{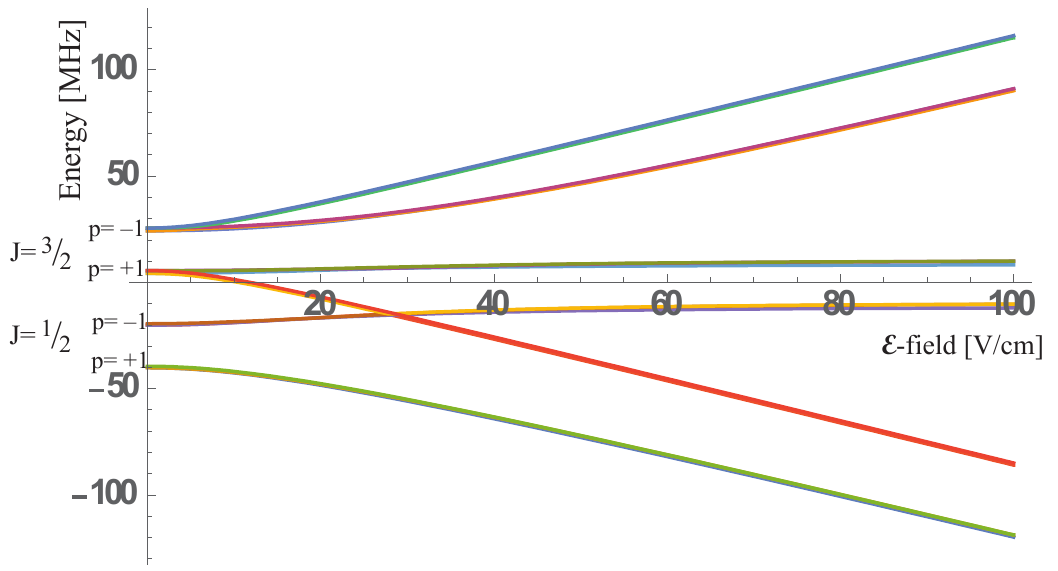}
	\caption{Stark shifts with $\gamma=$30 MHz, $q=-10$ MHz, $b=2$ MHz, and $d=4$ D, representative of YbOH.  There are several avoided and unavoided crossings.  The states are labeled on the left side by their zero-field quantum numbers.}
	\label{fig:StarkShifts}
\end{figure}

Consider the stretched states with $M=\pm 2$.  These are very simple to understand since there are only four states in this manifold ($M=\pm 2, P=\pm 1$, $J=3/2$, $F=2$), all matrix elements conserve $M$, and the Stark matrix elements are
\begin{multline}
\braket{P',J=3/2,F=2,M' | H_S | P,J=3/2,F=2,M} \\
= \begin{cases}
-1/2 & P' \neq P, M'=M \\
0 & \textrm{otherwise}
\end{cases}
\end{multline}
This means that the fully-polarized eigenstates are states of good $\pm M, \pm \ell$, i.e. fully mixed parity.  If we examine the eigenstates individually, we see that the sign of the Stark shift is $\propto -M\ell$, which is reminiscent of the Stark shift in a $\Omega>0$ diatomic state of $\propto -M\Omega$.

Table \ref{tab:StateComposition} shows the composition of the eigenstates in a 250 V/cm field in terms of the zero-field eigenstates.  There is an almost-symmetry between the top and bottom half of the table; flipping the parity in the table would yield nearly the same table (up to signs), except for a few states that have different relative admixtures of $F$ values in $\pm P$ (for example, the states in columns 2 and 3). These states also have the additional interesting feature that they have $M=0$ and therefore no linear Zeeman shifts but do have linear Stark shifts.

\subsubsection{Zeeman Shifts}

Now let's apply a magnetic field in the $z-$direction.  We shall again only consider $M\geq 0$, keeping in mind that the (relevant) linear Zeeman shifts are odd in $M$.  We will work under the conditions where the Zeeman shift is smaller than any other energy splitting, so we can treat the effect perturbatively.  We will also ignore any ``small" effects, such as nuclear spin, rotation, and quadratic Zeeman shifts, considering only linear magnetic interactions from the electron spin with the applied field.  The matrix elements are given by Hirota (2.5.16):
\begin{multline}
\braket{N'\ell'SJ'IF'M'|H_Z|N\ell SJIFM} = \\
\delta_{NN'}\delta_{\ell\ell'}(-1)^{F'-M}[(2F'+1)(2F+1)]^{1/2}\begin{pmatrix}F' & 1 & F \\ -M' & 0 & M\end{pmatrix} \\
\times (-1)^{F'+I+F+1}[(2J'+1)(2J+1)]^{1/2}\begin{Bmatrix}J' & F' & I \\ F & J & 1\end{Bmatrix} \\
 \times (-1)^{N+S+J'+1}[S(S+1)(2S+1)]^{1/2}\begin{Bmatrix}S & J' & N \\ J & S & 1\end{Bmatrix}
\end{multline}

We can find the Zeeman shifts in the polarized limit by taking the eigenstates in some $\E-$field and finding the expectation of the Zeeman operator.  The results are shown in table \ref{tab:StateComposition}.

\begin{widetext}

%\onecolumngrid

\scriptsize

\begin{table}[h]
\begin{tabular}{cccc|cccccccccccccccc}
$J$ & $F$ & $M$ & $P$ & & & & & & & & & & & & & & & &  \\ \hline
 1/2 & 0 & 0 & --1 & 0. & 0. & 35.3 & 0. & 0. & 1.3 & 0. & 0. & 0. & -33.3 & 0. & 0. & 0. & -30. & 0. & 0. \\
 1/2 & 1 & 0 & --1 & 0. & 35.3 & 0. & 0. & 0. & 0. & 1.6 & 0. & -33. & 0. & 0. & 0. & -30. & 0. & 0. & 0. \\
 1/2 & 1 & 1 & --1 & 35.3 & 0. & 0. & 0. & 0. & 0. & 0. & 1.5 & 0. & 0. & -33.2 & -29.9 & 0. & 0. & 0. & 0. \\
 3/2 & 1 & 0 & --1 & 0. & 12.6 & 0. & 0. & 0. & 0. & -0.1 & 0. & 65.3 & 0. & 0. & 0. & -22. & 0. & 0. & 0. \\
 3/2 & 1 & 1 & --1 & -3.9 & 0. & 0. & 35.3 & 0. & 0. & 0. & 0. & 0. & 0. & -16.3 & 6.6 & 0. & 0. & -37.9 & 0. \\
 3/2 & 2 & 0 & --1 & 0. & 0. & 12.6 & 0. & 0. & 0. & 0. & 0. & 0. & 65.3 & 0. & 0. & 0. & -22. & 0. & 0. \\
 3/2 & 2 & 1 & --1 & 8.8 & 0. & 0. & 12.7 & 0. & 0. & 0. & 0. & 0. & 0. & 48.9 & -15.5 & 0. & 0. & -14. & 0. \\
 3/2 & 2 & 2 & --1 & 0. & 0. & 0. & 0. & 48. & 0. & 0. & 0. & 0. & 0. & 0. & 0. & 0. & 0. & 0. & -52. \\
 1/2 & 0 & 0 & +1 & 0. & 38.6 & 0. & 0. & 0. & 0. & -33.4 & 0. & 0. & 0. & 0. & 0. & 28. & 0. & 0. & 0. \\
 1/2 & 1 & 0 & +1 & 0. & 0. & 38.6 & 0. & 0. & -33.3 & 0. & 0. & 0. & 0. & 0. & 0. & 0. & 28. & 0. & 0. \\
 1/2 & 1 & 1 & +1 & 38.5 & 0. & 0. & 0. & 0. & 0. & 0. & -33.4 & 0. & 0. & 0. & 27.9 & 0. & 0. & 0. & 0. \\
 3/2 & 1 & 0 & +1 & 0. & 0. & 13.4 & 0. & 0. & 65.3 & 0. & 0. & 0. & 1.3 & 0. & 0. & 0. & 19.9 & 0. & 0. \\
 3/2 & 1 & 1 & +1 & -4.1 & 0. & 0. & 38.3 & 0. & 0. & 0. & -16.2 & 0. & 0. & -0.4 & -6. & 0. & 0. & 35. & 0. \\
 3/2 & 2 & 0 & +1 & 0. & 13.4 & 0. & 0. & 0. & 0. & 64.9 & 0. & 1.7 & 0. & 0. & 0. & 20. & 0. & 0. & 0. \\
 3/2 & 2 & 1 & +1 & 9.4 & 0. & 0. & 13.7 & 0. & 0. & 0. & 48.8 & 0. & 0. & 1.1 & 14. & 0. & 0. & 12.9 & 0. \\
 3/2 & 2 & 2 & +1 & 0. & 0. & 0. & 0. & 52. & 0. & 0. & 0. & 0. & 0. & 0. & 0. & 0. & 0. & 0. & 48. \\  \hline
 \text{} & \text{} & \text{} & $\Delta_Z$ & -0.985 & 0. & 0. & 0.998 & 1. & 0. & 0. & 0.986 & 0. & 0. & 0.986 & -0.981 & 0. & 0. & 0.997 & 1. \\
 \text{} & \text{} & \text{} & $P_E$ & -0.993 & -0.993 & -0.993 & -0.999 & -0.999 & 0.001 & 0.001 & 0.001 & 0. & 0.001 & 0.001 & 0.991 & 0.991 & 0.991 & 0.999 & 0.999 \\
& & & $M$: & 1 & 0 & 0 & 1 & 2 & 0 & 0 & 1 & 0 & 0 & 1 & 1 & 0 & 0 & 1 & 2
\end{tabular}
\normalsize
\caption{Table of admixtures with a 250 V/cm electric field, assuming $\gamma=30$ MHz, $b=2$ MHz, $q=-10$ MHz, and $d$ = 4 D.  The values are fractional admixtures of amplitude (in percent), where the sign indicates the sign of the contribution before squaring.  At the bottom we list the signed electric polarization $P_E$, the Zeeman shift in units of $\mu_B\B$, and the value of $M$, which is conserved for our case.}
\label{tab:StateComposition}
\end{table}
\normalsize

\end{widetext}
%\twocolumngrid
\normalsize

\subsection{Estimate of Franck-Condon factors for polyatomics}

\begin{table}
\begin{centering}
\begin{tabular}{|c|c|c|}
\hline 
Transition & FCF & VBR\tabularnewline
\hline 
\hline 
\multicolumn{3}{|c|}{$\tilde{X}^{2}\Sigma^{+}-\tilde{A}^{2}\Pi_{1/2}$}\tabularnewline
\hline 
(000)-(000) & 0.8674 & 0.8777\tabularnewline
\hline 
(100)-(000) & 0.1174 & 0.1082\tabularnewline
\hline 
(200)-(000) & 0.0133 & 0.0111\tabularnewline
\hline 
(300)-(000) & 0.0013 & 0.0010\tabularnewline
\hline 
(020)-(000) & 0.0006 & 0.0006\tabularnewline
\hline 
(001)-(000) & $2\times10^{-5}$ & $1\times10^{-5}$\tabularnewline
\hline 
\end{tabular}
\par\end{centering}
\protect\caption{Estimated Franck-Condon factors (FCF) and vibrational branching ratios (VBR)
for the $\tilde{X}-\tilde{A}$ transition in \protect\textsuperscript{174}YbOH.
The calculation was performed assuming the harmonic oscillator approximation
of molecular vibrations using methods from Ref. \cite{sharp1964franck}. Measured
YbOH molecular constants from Ref. \cite{Melville2001} were used as input. Our
calculations were initially benchmarked by comparing measured and
calculated FCFs for SrOH \cite{KozyryevThesis} and showed close agreement.}
\label{tab:YbOH_FCF}
\end{table}

The FCFs for YbOH are shown in table \ref{tab:YbOH_FCF}.  Similar estimates for YbCCH,
YbOCH\textsubscript{3}, and YbCH\textsubscript{3} indicate that two dominant FCFs for Yb-ligand stretching vibrations sum to $\gtrsim0.95$ for all of these species, making them all promising candidates for laser cooling. While including anharmonic terms in the molecular
potential will improve the accuracy of the calculation, the relative
magnitudes of the loss channels as well as the total sum of the dominant
FCFs should not change significantly as indicated by our SrOH studies \cite{KozyryevThesis}. Our calculated FCFs for YbOH are comparable to those measured for the isoelectronic diatomic molecule YbF \cite{Zhuang2011}.

\subsection{Populating the excited vibrational state}

\begin{figure}
\begin{centering}
\includegraphics[width=7cm]{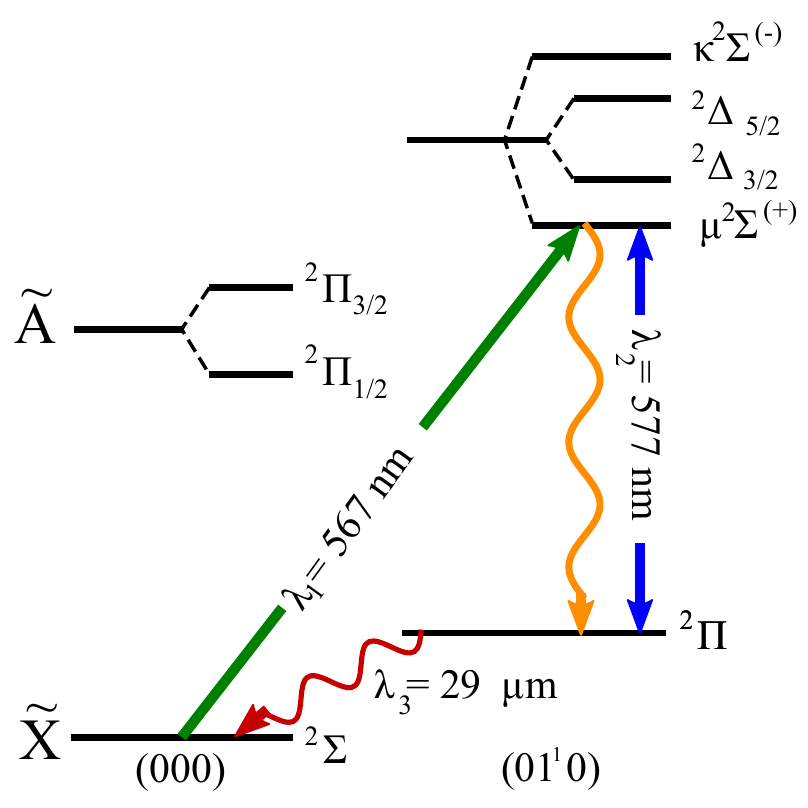}\protect\caption{\label{fig:Proposed-optical-pumping}Proposed optical pumping scheme for YbOH in order to populate the excited bending state (010). Vibronic species of the relevant states are indicated. Optical readout of the molecular population in (010) state can be performed via direct excitation with $\lambda_2$.  The $\tilde{A}$ state splits into four distinct states due to the spin-orbit and Renner-Teller interactions. The energy splittings are not to scale. }
\par\end{centering}
\end{figure}

Even though $\triangle l_{2}\neq0$ vibronic transitions are forbidden
in the Born-Oppenheimer (BO) approximation, optical excitation of the nominally forbidden
$\tilde{X}\left(000\right)\rightarrow\tilde{A}\left(010\right)$ transition
has been previously observed for alkaline earth monohydroxides. The spin-orbit
(SO) vibronic Renner-Teller couplings $H_{RT}\times H_{SO}$ mix $\tilde{A}^{2}\Sigma^{+}$
and $\tilde{B}^{2}\Pi$ states with $v_{2}=\pm1$ and $\triangle l_{2}=-\triangle\Lambda=\pm1$
resulting in the BO approximation breakdown \cite{presunka1994laser}. Assuming comparable size
of the Renner-Teller parameter $\epsilon$, the magnitude of the forbidden
transition probability should scale as $A_{SO}^{2}/\triangle E_{\Sigma-\Pi}^{2}$
which is approximately the same value for BaOH and YbOH. Previous
experimental measurements for BaOH \cite{kinsey1986rotational} show comparable strength
of the allowed $(02^{0}0)-(000)$ and forbidden $\left(010\right)-\left(000\right)$
bands. Efficient optical pumping into the excited bending
state $\tilde{X}\left(010\right)$ via $\tilde{X}^{2}\Sigma^{+}\left(000\right)\rightarrow\mu\tilde{A}^{2}\Sigma^{\left(+\right)}$ off-diagonal
excitation should be possible as previously experimentally seen for
SrOH \cite{presunka1994laser}. Figure \ref{fig:Proposed-optical-pumping} depicts the optical pumping scheme that can be used for transferring YbOH molecules from the vibrational ground state to the (010) excited bending mode where the T-violating effects will be measured.   Alternatively, optical pumping can be replaced with coherent Raman transfer \cite{panda2016stimulated} leading to greater efficiency during the state transfer process.

\subsection{Estimate of vibrational lifetimes}

To estimate the lifetime of the YbOH $\nu_2=1$ excited bending mode, we use the measured excitation energy $\sim300$ cm$^{-1}$ and estimated (000)-(010) transition dipole moment of $0.1$ D \cite{Okumura1987,senekowitsch1987vibrational} to compute an Einstein A-coefficient of $\sim 0.1$ s$^{-1}$, or a lifetime of $\gtrsim 10$ s.  Black-body radiation (BBR) can also induce vibrational decays, but we estimate that this effect at 300 K should be lower than the spontaneous decay rate by $\sim$1/4 and therefore will not limit the experimental coherence time. Additionally, reducing the chamber temperature to 77 K will reduce $\Gamma_{\rm{BBR}}$ by a factor of 400. Our estimations are consistent with the previous analysis of black-body limited vibrational lifetimes for diatomic molecules \cite{vanhaecke2007precision,buhmann2008surface}.

\subsection{Laser cooling with multiple ground states}

Coupling multiple ground states to a few excited states results in
reduction of the effective scattering rate and, correspondingly, decreased
radiative force \cite{Tarbutt2013}. Decoupling of the main cycling transition $\lambda_{0}$
from multiple repumping lasers will lead to rapid optical cycling
at multiple MHz rates. Spectroscopy of the isoelectronic molecule
YbF indicates the presence of the $\tilde{B}^{2}\Sigma^{+}$ molecular
state originating from the $4f^{14}6p\sigma$ atomic orbital of the
Yb\textsuperscript{+} ion \cite{barrow1975analysis}. Repumping the (020), (200), and
(300) excited vibrational levels through the $\tilde{B}$ state
will increase the cycling rate by a factor of 2.3 leading to stronger
radiative force. Moreover, promotion of the $f$ electrons into
the unfilled excited orbitals leads to additional electronic states
not present in alkaline earth monohydroxides. Repumping through
such levels has been considered possible for YbF \cite{smallman2014radiative} and should work for YbOH as well for repumping dark vibrational states which are infrequently
populated (e.g. (300)). 

The use of coherent stimulated optical forces instead of traditional radiative techniques for beam deceleration and cooling will significantly reduce the number of required spontaneous emissions. Particularly, the bichromatic force has been extensively investigated theoretically for complex multilevel molecules \cite{aldridge2016simulations} and recently demonstrated experimentally for a triatomic molecule SrOH \cite{KozyryevThesis}.  

\subsection{Symmetric tops}

Since the energy of symmetric top molecules  is the same for clockwise and counterclockwise rotations around the top symmetry axis, the states with $\pm K$ should have the same energy, where $K$ is the projection of the total rotational, orbital, and vibrational angular momentum on the symmetry axis. Since the ground electronic and vibrational levels of molecules isoelectronic to YbOH like YbCH\textsubscript{3} and YbOCH\textsubscript{3} have no orbital or vibrational angular momenta, $K$ solely represents the projection of rotational angular momentum $R$.  The presence of any interaction that directly or indirectly couples the $\pm K$ results in the eigenstates $\lvert R,\,+K\rangle\pm\lvert R,\,-K\rangle$,  which represent doublets of opposite parity. Like $\ell$-doublets for linear triatomics in excited bending levels, such $K$-type doublets arise from a slight molecular asymmetry but the doublet splitting is typically much smaller. Previous studies have shown that for XCH$_3$ type molecules, the largest doublet splitting is for $\lvert K\rvert=1$ levels and arises from H spin-rotation and X-H spin-spin coupling resulting in kHz-wide splittings. Splittings between $K-$doublets for $K\geq2$ are significantly smaller, with about $10^{-4}\,$ Hz for $K=2$ \cite{klemperer1993can}. Consequently, complete polarization of symmetric top molecules is possible for rotational states with $K\neq0$ in very small laboratory electric fields of $\sim1\,$mV/cm \cite{klemperer1993can,butcher1993hyperfine} which is even smaller than for $\ell-$doublets. 

The lowest rotational level of each $K$ has energy $\approx AK^2$. Therefore, compared to the $\sim10\,$THz excitation energies of the lowest bending mode levels with non-zero $\ell$ of YbOH, the excited $K=1$ states with opposite parity $K$ doublets are typically only $\sim160\,$GHz above the absolute vibronic ground state for symmetric top YbCH$_3$ and YbOCH$_3$. Thus, the spontaneous decay lifetimes are typically well over one minute. Since ortho and para configurations of the CH$_3$ group of XCH$_3$ and XYCH$_3$ molecules do not efficiently cool into each other in the collisions associated with supersonic expansion cooling \cite{dick2006high}, we anticipate that a large fraction of molecules in $K=1$ rotational levels will also be created during the buffer-gas cooling process as previously seen for some symmetric tops \cite{wu2016thermometry}. Alternatively, direct optical pumping into excited $K$ levels is possible by using perpendicular optical transitions with $\triangle K=\pm1$ selection rule (e.g. $\tilde{X}-\tilde{A}$ band for SrCH$_3$ and SrOCH$_3$ \cite{dick2006high}). The cylindrical symmetry of symmetric tops enables well-defined rotational selection rules that can be used to achieve optical cycling on a quasi-closed transition with only a few laser frequencies. The specific details of achieving optical cycling in alkaline earth monoalkoxides MOR like CaOCH$_3$ have been laid out in detail in Ref. \cite{Kozyryev2016Poly} and because of their electronic structure similarity with YbOR, the same approach is applicable here as well. Additionally, the dominant four Franck-Condon factors for YbOCH$_3$ sum up to $>0.99$, indicating that efficient optical cycling and laser cooling could be achieved. 

\bibliography{library}

\end{document}